\documentclass{article}
\usepackage[utf8]{inputenc}
\usepackage{amsmath}
\usepackage{relsize}
\usepackage{url}
\usepackage{graphicx}

\def\code#1{\textsc{#1}}
\usepackage{color}
\definecolor{dkgreen}{rgb}{0,0.6,0}
\definecolor{gray}{rgb}{0.5,0.5,0.5}
\definecolor{mauve}{rgb}{0.58,0,0.82}

\definecolor{dkgreen}{rgb}{0,0.6,0}
\definecolor{dkblue}{rgb}{0,0,0.6}
\definecolor{gray}{rgb}{0.5,0.5,0.5}
\definecolor{mauve}{rgb}{0.58,0,0.82}

\usepackage{xcolor}
\definecolor{commentgreen}{RGB}{2,112,10}
\definecolor{eminence}{RGB}{108,48,130}
\definecolor{weborange}{RGB}{255,165,0}
\definecolor{frenchplum}{RGB}{129,20,83}

\usepackage{listings}
\title{A Stack-Free Traversal Algorithm for Left-Balanced k-d Trees}
\author{Ingo Wald}

\date{(revision 1)}

\begin{document}

\maketitle
  \vspace{-2em}
\begin{abstract}
  We present an algorithm that allows for find-closest-point and
  kNN-style traversals of left-balanced k-d
  trees, without the need for either recursion or software-managed
  stacks; instead using only current and last previously traversed node
  to compute which node to traverse next.
\end{abstract}

\section{Introduction}

\begin{figure}[ht]
  \vspace{-1em}
  \centering
  \begin{tabular}{cc}
    \begin{minipage}{.48\textwidth}
      \vspace*{-5cm}
      \includegraphics[width=.99\textwidth]{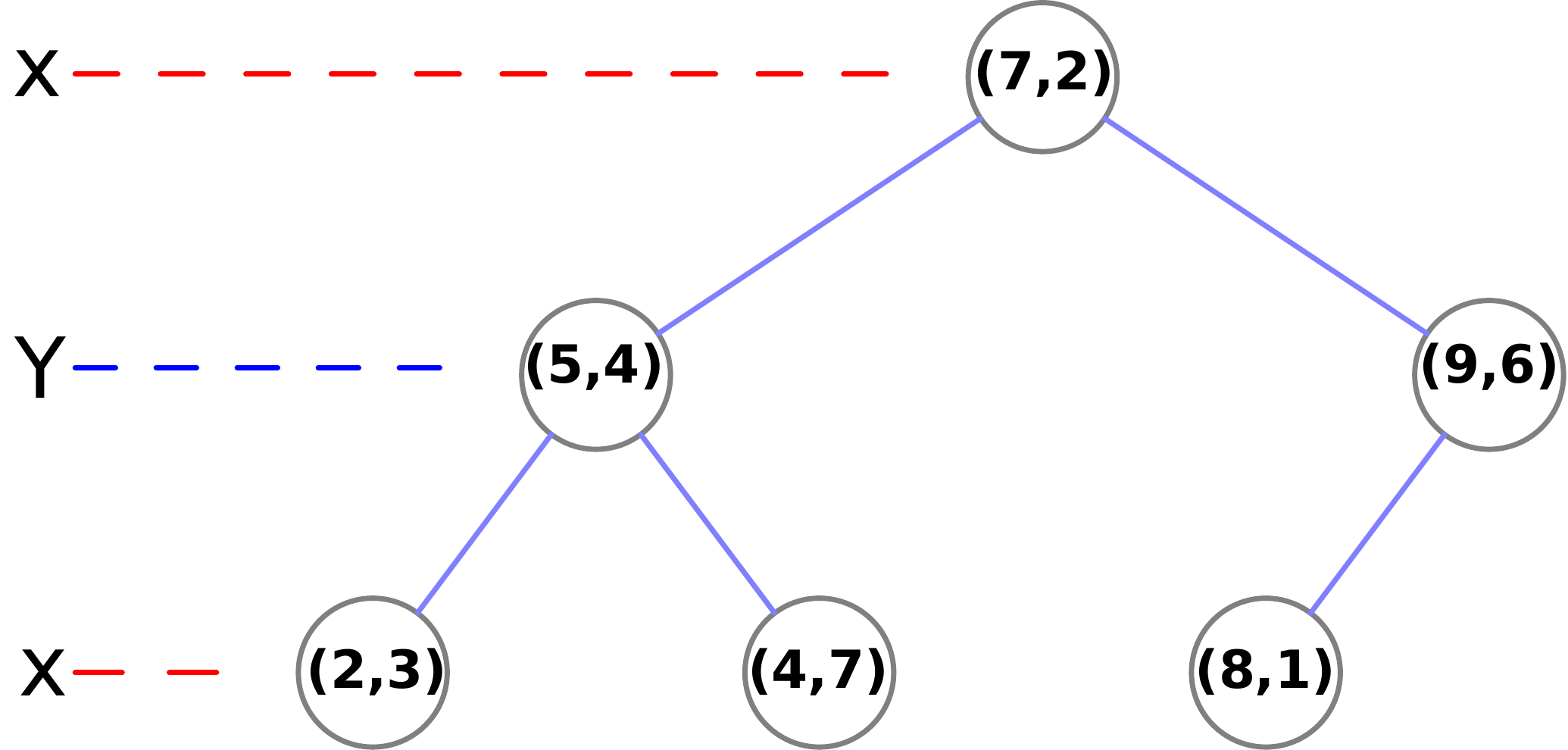}
    \end{minipage}
    &
    \includegraphics[width=.4\textwidth]{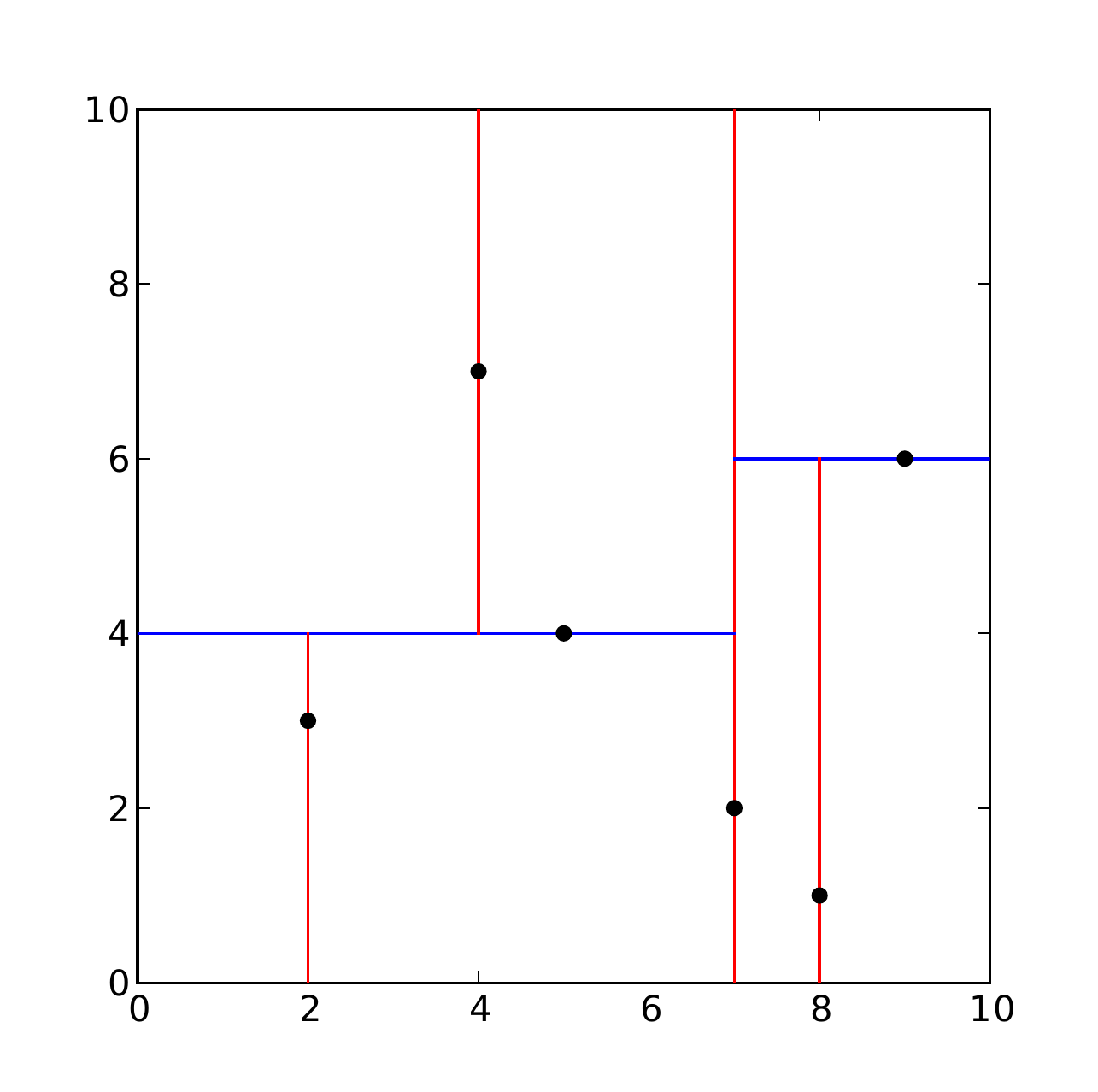}
  \end{tabular}
  \vspace{-2em}
  \caption{\label{fig:kd} Illustration of a 2-dimensional k-d tree
    from Wikipedia~\cite{wiki-kdtree}.
    Left: The balanced tree for
    2D point
    set $\{(2,3), (5,4), (9,6), (4,7), (8,1), (7,2)\}$.
    Right: The space partitioning achieved by that tree. }
\end{figure}

K-d trees (see Figure~\ref{fig:kd})
are powerful, versatile, and widely used spatial data
structures for storing, managing, and performing queries on
k-dimensional data.  One particularly interesting
type of k-d trees are those that are left-balanced and complete: for
those, storing the tree's nodes in level order means that the entire
tree topology---i.e., which are the parent, left, or right child of a
given node---can be deduced from each node's array index. This means
that such trees require zero overhead for storing child pointers or
similar explicit tree topology data, which makes them particularly
useful for large data, and for devices where memory is precious (such
as GPUs, FPGAs, etc).
Throughout the rest of this
paper we will omit the adjectives left-balanced and complete, and
simply refer to "k-d trees"; but always mean those that are both
left-balanced and complete.

One issue with k-d trees is that their hierarchical nature means that
they work best with recursively formulated traversal techniques; but
recursive formulations can be problematic on highly parallel
architectures such as, for example, GPUs, FPGAs, dedicated hardware
units, etc: true recursion---where functions can call themselves---is
often hard or impossible to realize on such architectures, and is
instead typically replaced with iterative algorithms that emulate
recursion through a manually managed stack of not-yet-traversed
sub-trees. Such ``manual'' stack based techniques do indeed avoid
truly recursive function calls---and thus work just fine on GPUs---but
can still cause several issues: for example, maintaining a stack per
live thread (of which there will typically be many thousands) can
require large amounts of temporary memory; it can lead to a
significant amount of memory accesses (since such stacks are
register-indirectly accessed they will typically end up in device
memory); they can lead to hard-to-track errors if the reserved stack
size isn't large enough; etc.

In this paper, we describe a stack-free traversal algorithm for
left-balanced k-d trees that does not require any stack at all, and
instead only requires two integer values (for current and last
traversed node IDs, respectively) to track all traversal state. In
particular, our algorithm can handle both simple unordered traversals
as well as the more channeling ordered kind where the traversal order
of two children depends on which side of the parent's partitioning plane
the query point is located on. Our algorithm borrows from ideas we had previously
developed for bounding volume hierarchies~\cite{hapala}, and relies on
using a \emph{state machine}-like approach that, in each step, derives
the respectively \emph{next} node to be traversed from some relatively
simple logic that only needs to know the current node that is currently traversed, and the
one that was traversed in the previous step. We describe the core algorithm,
provide some simple pseudo-code that realizes it, and present some
performance data for a CUDA implementation of various variations of
\emph{find-closest-point} (\code{fcp}) and \emph{k-nearest-neighbor}
(\code{kNN}) queries that use this algorithm.

\section{Related Work}

k-d Trees (sometimes also referred to as kd-trees) are hierarchical
spatial search trees used to encode k-dimensional data
points~\cite{samet1990design,knuth1997art,wiki-kdtree}. Though the same term in
graphics is also sometimes used to refer to a different kind of
spatial subdivision tree (where inner code describe arbitrary
partitioning planes and data are only stored in the
leaves~\cite{Wald_building_2006} ) for this paper we explicitly only
consider the former kind, in which the data to be encoded are
k-dimensional points, where points are stored also at inner nodes, and
where partitioning planes are defined \emph{by} the coordinates of the
points stored at inner nodes (i.e., partitioning planes always pass
\emph{through} data points; see Figure~\ref{fig:kd}).

Though k-d trees do not necessarily \emph{have} to be balanced, one
particularly interesting type of k-d trees are those that are
left-balanced and complete, since these allow for storing the tree
without any pointers or other explicit topology data. In graphics,
these became famous through \emph{photon
  mapping}~\cite{jensen2001realistic}, in which k-d trees were used to
memory-efficiently store---and efficiently perform k-nearest neighbor
queries on---large numbers of photon. They have since become a method
of choice for any kind of technique that requires storing and querying
large numbers of point data.

This paper is about \emph{traversing} such k-d trees, and in
particular, in doing that in a stack-free manner. Mainly caused by the
desired to finally achieve real-time ray tracing on graphics
processing units (GPUs) the last decade has seen a concerted effort to
develop such stack-free traversal techniques for ray traversal through
bounding volume hierarchies
(e.g.,~\cite{laine2010restart,hapala,barringer2013dynamic,afra2014stackless,binder2016efficient,vaidyanathan2019wide}). Such
BVHes are different from k-d trees (and the queries we are targeting
are different from ray traversal) but the motivation in both cases is the
same: maintaining stacks can cause issues in highly parallel
architectures where any per-thread state is potentially costly.
In this paper, we primarily build on the technique described
in~\cite{hapala}, but adapt that to k-d trees and \code{fcp}/\code{kNN}-style traversals.

\section{Algorithm}

One issue with traversal algorithms for hierarchical data structures
is that their requirements often come in many similar yet nevertheless slightly different
forms; for example, whether the traversal need to be ordered vs
un-ordered (i.e., whether the traversal order of two child nodes
depends on the given query primitive or not); or whether the
partitioning dimension of a k-d tree node is explicitly stored with the
node vs implicitly fixed by the level of the node in the tree; or what
exactly the traversal does with each visited data point (e.g.,
find-closest-point \code{fcp} vs k-nearest neighbors \code{kNN}); or whether the
query primitive is a point/sphere (to find closest point or kNNs to) vs a box
(to find all points within this box); etc.

In this section, we describe our algorithm for the following configuration:
\begin{itemize}
  \item we assume an \emph{ordered traversal}, where the traversal
    order for two children should always first traverse the child that
    is on the same side of the partitioning plane as the query point;
    a non-ordered traversal is a trivial simplification of this case.
  \item since ordered traversals are typically ordered relative to a
    single query point we assume a \emph{point} as query primitive;
    replacing that with different query primitive types should be
    trivial.
  \item we assume that there is a \emph{shrinking max radius} for the
    query, where the traversal can, while traversing, decide to
    exclude ever larger portions of the full tree based on whatever
    partial solution has already been found.
  \item we assume the tree is stored in the canonical way of using
    level-order (i.e., the root node is stored at array index 0, and
    the left and right child of a node $n$ are stored at
    $lChild(n)=2n+1$ and $rChild(n)=2n+2$).
  \item we make no assumptions about what partitioning plane dimension
    the builder chose for any given node; we simply assume existence of a
    function \code{splitDimOf(node)} that returns the partitioning
    dimension of this node. 
\end{itemize}

\subsection{Deriving the Traversal Logic}

Using these assumptions, the core idea of our algorithm is to use a
``state machine'' like approach where we only track two integer
values---the current node to be traversed, and the one that was
traversed in the previous step---and use some simple observation to
derive the next to-be-traversed node from that information. We observe
that in a recursive formulation, the ordered traversal for any given
sub-tree in a k-d tree looks like this\footnote{This formulation
  assumes that the root node of any sub-tree always gets processed
  \emph{before} the closer child sub-tree. Past experience leads
  us to believe this to usually be the preferred order in particular
  for shrinking-radius queries, but changing the order of root and close
  sub-tree in the logic below is possible, too, and straightforward.}:
\begin{enumerate}
\item process the current node (possibly shrinking the max
  query radius).
\item compute which of the two sub-trees is closer to the query point.
\item if close child exists: (recursively) traverse the sub-tree of the closer child.
\item if far child exists \emph{and} sub-tree of far child still overlaps the current query range,
  recursively traverse that sub-tree.
\end{enumerate}

\medskip\noindent For simplicity, let us re-formulate this as
\begin{enumerate}
\item if sub-tree is empty, return without doing anything.
\item process the current node (possibly shrinking the max query
  radius).
\item compute which of the two sub-trees is closer to the query point.
\item (recursively) traverse the sub-tree of the closer child.
\item if sub-tree of far child still overlaps the current query range,
  recursively traverse that sub-tree.
\end{enumerate}
This is the exact same logic, except with the ``if child exists''
moved into the code for processing a given node.

If we now only look at any one particular node, and consider all the
different cases that this recursive traversal would ever be ``at'' a
given node, we can identify the following cases:
\begin{enumerate}
\item if the node in question does not exist, traversal will not do
  anything, and will simply go back to the parent node.
\item the \emph{first} time a (valid) node gets traversed is when its
  traversal gets first called from its parent; in this case, it will
  first process the current node, and then go to the \emph{close}
  child (which may or may not actually exist).
\item the \emph{second} time a node gets visited is when recursion
  returns from the close child; in this case, there are two possible
  cases where the traversal will go next: if the far sub-tree is
  still in range, the next node to be traversed is the
  \emph{far} child; otherwise, traversal is returning and the next
  node to traverse is the parent.
\item the \emph{third} (and last) possible case where a node can gets
  visited is when recursion returns from the far child; in this case
  the next node to go to is the parent.
\end{enumerate}
Using these rules, all we need to compute the next node to traverse is
knowing which of these three cases we are in at any given point in
time.  The first case can be detected trivially by simply checking if
the current node index is valid; the other three we can detect simply
based on what the \emph{last} traversed node was:
\begin{itemize}
\item the first-time-visit case only gets reached when the previous node
  was the parent, so if the previous node was the parent we process
  the current node and go to the close child.
\item the second-time-visit case can only be reached if the previous node
  was the close child; in this case we need to test if the far sub-tree
  is in range, and next go to either far child or parent.
\item the third-time-visit case can only be reached if the previous node
  was the far child; in this case we go to the parent (or terminate if
  we are at the root, obviously).
\end{itemize}
Using this formulation and applying this to every node that we are
traversing will necessarily iterate through exactly the same
nodes---in exactly the same order---as the recursive formulation we
used above. All we need to do is keep track of both current and
previous node ID, and apply this logic.

\subsection{Pseudo-Code}

In CUDA-like pseudo-code the algorithm can be expressed as follows:
\begin{lstlisting}
void traverse(point queryPos, node *nodes, int N) 
{
  // current node: start at root
  int curr = 0;
  // previous node, initialize to "parent" of root node
  int prev = -1;
  float maxSearchRadius = ...;
  // repeat until we break out:
  while (true) {
     int parent = (curr+1)/2-1;
     if (curr >= N) {
       // we reached a child that does not exist; go back to parent
       prev = curr; curr = parent; continue;
     }
     bool from_parent = (prev < curr);
     if (from_parent) { 
        processNode(curr);
        // check if processing current node has led to 
        // a smaller search radius:
        maxSearchRadius = possiblyShrunkenSearchRadius();
     }
     // compute close and far child:
     int   splitDim   = splitDimOf(nodes,curr);
     float splitPos   = nodes[curr].coords[splitDim];
     float signedDist = queryPoint[splitDim] - splitPos;
     int   closeSide  = (signedDist > 0.f);
     int   closeChild = 2*curr+1+closeSide;
     int   farChild   = 2*curr+2-closeSide;
     bool  farInRange = (fabsf(signedDist) <= maxSearchRadius);

     // compute next node to step to
     int next;
     if (from_parent) 
       next = closeChild;
     else if (prev == closeChild)
       next = (farInRange ? farChild : parent);
     else {
       next = parent;
     }
     if (next == -1)
       // the only way this can happen is if the entire tree under
       // node number 0 (i.e., the entire tree) is done traversing,
       // and the root node tries to step to its parent ... in
       // which case we have traversed the entire tree and are done.
       return;
     // aaaand ... do the step
     prev = curr; curr = next;
  }
}
\end{lstlisting}

\section{Evaluation}

A sample implementation of this algorithm for various configurations
of \emph{find-closest-point} (\code{fcp}) and
\emph{k-nearest-neighbors} (\code{kNN}) can be found on github, at
\code{https://github.com/ingowald/cudaKDTree}
(see~\cite{github-repo}).  This repository also contains a gpu-based
parallel k-d tree builder that is described in a related
publication~\cite{kd-construction}, but which this paper will not
further comment upon other than observing that any tree built with
this method will be the same exact left-balanced k-d tree that any
other serial or CPU-based builder for such trees would produce, too.

Our sample traversal code is currently hard-coded for only two
kernels: \code{fcp}, and \code{knn} for various $k$; but should be
easy to adapted to other queries as well (in fact, it can easily be
templated over a \code{processNode()} lambda). Similarly this code is
hard-coded to \code{float4} data types, but can trivially be adapted
to (or templated over) other point types, including those with
additional payload data (like, for example, surface normal and photon
power for  photon mapping), or to node types with an explicitly stored
split dimension, etc.

To at least briefly evaluate this algorithm, we also added a very
simple test rig that generates two arrays of uniformly distributed
random \code{float4} points in $[0,1]^4$, using a user-specified
number of $N$ data points and a fixed $M=10,000,000$ query points;
this test rig then then performs queries like \code{fcp} and
\code{kNN} for various $k$ (including bounded and bounded variants for
the \code{kNN} case). We run these on an NVIDIA RTX~3090TI card with
10,752 CUDA cores and 1.86~GHz boost clock, using Ubuntu 20.04.5~LTS,
driver 510.85, and CUDA version 11.4; the results of these experiments
are given in Table~\ref{tab:results}.
\begin{table}[h!]
  \begin{center}
    {\relsize{-1}{
\begin{tabular}{c||c|cccc|cccc}
  N & \code{fcp}
  & \multicolumn{4}{c}{\code{knn} with k=... and maxR=$\inf$}
  & \multicolumn{4}{c}{\code{knn} with k=... and maxR=0.01}
  \\
  &
  & k=4 & k=8 & k=20 & k=50
  & k=4 & k=8 & k=20 & k=50
  \\
    \hline
    \hline
    1k   & 888M &  563M & 359M  & 48M   &   14M &  5.2G &  6.6G &  4.0G & 1.8G \\
    10k  & 664M &  407M & 262M  & 29.5M &  8.7M &  3.3G &  3.9G &  3.3G & 1.6G \\
    100k & 396M &  184M & 122M  & 20.6M &  6.0M &  1.7G &  1.8G &  1.7G & 1.1G \\
    1M   & 100M & 42.1M & 26.6M & 7.5M  &  2.9M &  334M &  338M &  282M & 255M \\
    10M  &  44M & 19.3M & 12.3M & 5.0M  &  2.3M & 58.2M & 58.5M & 52.8M & 51M \\
    100M &  36M & 16.5M & 10.8M & 4.7M  &  2.1M & 19.1M & 16.6M & 14.2M & 13.8M \\
    200M & 34.8M & 16.2M & 10.6M & 4.5M & 2.1M & 16.7M & 12.2M & 9.2M& 8.9M \\
\end{tabular}
}}
  \end{center}
  \vspace{-2ex}
\caption{\label{tab:results} Results for our sample k-nearest neighbor
  (\code{knn}) and find-closest-point (\code{fcp}) use cases of our
  stack-free traversal algorithm, for various problem sizes and
  configurations. All order-of-magnitude indicators are multiples of
  1000, not 1024, so ``1M'' means ``1,000,000''. All results measured
  on an NVIDIA RTX~3090TI card, using the given number of (uniformly
  distributed) random input points in $[0,1]^4$, with 10M also uniform
  random distributed query points per kernel launch, and averaged over
  100 such launches.}
\end{table}

\section{Summary and Conclusion}

In this paper, we have described a stack-free k-d tree traversal
algorithm that works on left-balanced k-d trees over k-dimensional
points, and that can do both ordered and non-ordered traversal without
requiring either recursion or manually managed stack; instead, all our
algorithm requires as per-thread state are two \code{int}s to track the
current and the respectively last previously traversed nodes, which
our algorithm then uses to determine what to do in the current node,
and where to go next.  Our algorithm is both general and flexible, and
on an RTX~3090TI and \code{float4} data (with random uniform
distributions) achieves traversal rates in the many millions of queries
per second for both \code{fcp} and \code{knn}.

We make no claims that our sample implementation be the fastest
possible implementation of this algorithm; nor even that this
algorithm (even if it was implemented optimally) was guaranteed to be
faster than a stack-based one; however, if and where a stack-free
implementation is either required or desired we believe this algorithm
to be a useful tool and viable solution.

\bibliographystyle{alpha}
\relsize{-1}{
  \bibliography{references} 
  }

\appendix
\section{Revision History}

\begin{description}
\item[Revision 1 (ArXiv version v2)]:
  \begin{itemize}
  \item Fixed two bugs in traversal
    pseudo-code that had crept in when simplifying the actual reference
    code into latex pseudo-code (original code and all measurements use
    correct code, only PDF had the bugs).
  \item Added cite and pointer to related parallel k-d tree construction paper
  \item Fixed missing URLs in References.
  \end{itemize}
\item[Original Version (ArXiv version v1)]: Originally uploaded to ArXiv Oct 23,
  2022 \url{https://arxiv.org/abs/2210.12859}
\end{description}

\end{document}